\documentclass{article}
\usepackage{amsfonts, amsmath, amssymb}

\begin{document}

\begin{center}
{\Large Codes over subsets of algebras obtained by the Cayley-Dickson process%
}%
\begin{equation*}
\end{equation*}

Cristina FLAUT%
\begin{equation*}
\end{equation*}
\end{center}

\textbf{Abstract. }{\small In this paper, we define binary block codes over
subsets of real algebras obtained by the Cayley-Dickson process and we
provide an algorithm to obtain codes with a better rate. This algorithm
offers more flexibility than other methods known until now, similar to
Lenstra's algorithm on elliptic curves compared with }$p-1${\small \
Pollard's algorithm.}

\medskip 



\begin{equation*}
\end{equation*}

1. \textbf{Introduction}%
\medskip

\ Integer Codes are codes defined over finite rings of integers modulo $%
m,m\in \mathbb{Z}.$ Since these codes have a low encoding and decoding
complexity, they had a significant development over the last years and are
suitable for application in communication systems (see [Ko, Mo, Ii, Ha, Ma;
10]).

Some other codes similar to the Integer Codes, such as for example codes
over Gaussian integers ([Hu; 94], [Gh, Fr; 10], [Ne, In, Fa, Pa; 01], [Ri;
95]) or codes over Eisenstein--Jacobi integers, have been intensively
studied in recent years.

In this paper, we will extend the study of Integer Codes to codes over
subsets of real algebras obtained by the Cayley-Dickson process. This idea
comes in a natural way, starting from same ideas developed by Hubner in [Hu;
94], in which \ he \ regarded a finite field as a residue field of the
Gaussian integer ring modulo a Gaussian prime, ideas extended to Hurwitz
integers in [Gu; 13] and to a subset of the Octonions integers in [Fl; 15].
In this way, we regard a finite field as a residue field modulo a prime
element from $\mathbb{V},$ where $\mathbb{V}$ is a subset of an algebra $%
\mathbb{A}_{t}\left( \mathbb{R}\right) ,$ where $\mathbb{A}_{t}\left( 
\mathbb{R}\right) $ is a real algebra$\ $obtained by the Cayley-Dickson
process and $\mathbb{V}$ has a commutative and associative ring structure.
In this way, we obtain an algorithm, called Main Algorithm, which allows us
to find codes with a good rate. This algorithm offers more flexibility than
other methods known until now, similar to Lenstra's algorithm on elliptic
curves compared with $p-1$\ Pollard's algorithm. It is well known that for a
prime $p,$ Lenstra's algorithm replace the group $\mathbb{Z}_{p}^{\ast }$
with the group of the rational points of an elliptic curve $\mathcal{C}_{1}$
over $\mathbb{Z}_{p}$ and, if \ this algorithm failed, the curve will be
replaced with another curve $\mathcal{C}_{2}$ over $\mathbb{Z}_{p}$ and we
can retake the algorithm (see [Si, Ta; 92]).

In the case of Main Algorithm, the algebra $\mathbb{A}_{t}\left( \mathbb{R}%
\right) $ and $w$ offer this kind of flexibility since, for the same prime $%
p,$ these can be changed and the algorithm can be retake.\medskip

\begin{equation*}
\end{equation*}

\textbf{2. Preliminaries}%
\begin{equation*}
\end{equation*}

In the following, we shortly recall the Cayley-Dickson process for the real
algebras. Let $\mathbb{A}$ be a finite dimensional unitary algebra over the
real field $\ \mathbb{R}$ with a \textit{scalar} \textit{involution }$\,\,%
\overline{\phantom{xx}}:\mathbb{A}\rightarrow \mathbb{A},\quad a\rightarrow 
\overline{a},$ $\,\,$ i.\thinspace e. a linear map satisfying the following
relations:$\,\,\,\,\,\overline{ab}=\overline{b}\overline{a},\quad \overline{%
\overline{a}}=a,$ and \ $a+\overline{a},a\overline{a}\in \mathbb{R}\ $for
all $a,b\in \mathbb{A}.$ The element $\,\overline{a}$ is called the \textit{%
conjugate} of the element $a,$ the linear form$\,\,\,\,\mathbf{t}:\mathbb{A}%
\rightarrow \mathbb{R},\quad \mathbf{t}\left( a\right) =a+\overline{a}$ and
the quadratic form $\mathbf{n}:\mathbb{A}\rightarrow \mathbb{R},\quad 
\mathbf{n}\left( a\right) =a\overline{a}$ are called the \textit{trace} and
the \textit{norm \ }of \ the element $a.$ Since $\mathbf{n}\left( a\right) =a%
\overline{a}=a\left( \mathbf{t}\left( a\right) -a\right) ,$ it results that $%
a^{2}-\mathbf{t}\left( a\right) a+\mathbf{n}\left( a\right) =0,$ for all
elements $a\in \mathbb{A},$ therefore $\mathbb{A}$ is a quadratic algebra.

Let $\gamma \in \mathbb{R}$ \thinspace be a fixed non-zero element. On the
vector space $A\oplus A,$ we define the following algebra multiplication 
\begin{equation*}
\left( a_{1},a_{2}\right) \left( b_{1},b_{2}\right) =\left(
a_{1}b_{1}+\gamma b_{2}\overline{a_{2}},\overline{a_{1}}b_{2}+b_{1}a_{2}%
\right) .
\end{equation*}%
We obtain an algebra structure over $\mathbb{A}\oplus \mathbb{A},$ denoted
by $\left( \mathbb{A},\gamma \right) $ and called the \textit{algebra
obtained from } $\mathbb{A}$ \textit{\ by the Cayley-Dickson process.} We
have $\dim \left( \mathbb{A},\gamma \right) =2\dim \mathbb{A}$.

Let $x\in \left( \mathbb{A},\gamma \right) $, $x=\left( a_{1},a_{2}\right) $%
. The map $\,$%
\begin{equation*}
\,\,\overline{\phantom{x}}:\left( \mathbb{A},\gamma \right) \rightarrow
\left( \mathbb{A},\gamma \right) \,,\,\,x\rightarrow \bar{x}\,=\left( 
\overline{a}_{1},-a_{2}\right) ,
\end{equation*}
is a scalar involution of the algebra $\left( \mathbb{A},\gamma \right) $,
extending the involution $\overline{\phantom{x}}\,\,\,$of the algebra $%
\mathbb{A}.$

\thinspace If we take $\mathbb{A}=\mathbb{R}$ \thinspace and we apply this
process $t$ times, $t\geq 1,\,\,$we obtain an algebra over $\mathbb{R},\,\,$%
\begin{equation*}
\mathbb{A}_{t}\left( \mathbb{R}\right) =\left( \frac{\gamma _{1},...,\gamma
_{t}}{\mathbb{R}}\right) .
\end{equation*}%
In this algebra, the set $\{e_{0}=1,e_{1},...,e_{n-1}\},n=2^{t},$ generates
a basis with the properties: 
\begin{equation*}
e_{i}^{2}=\gamma _{i}1,\,\,\gamma _{i}\in \mathbb{R},\gamma _{i}\neq
0,\,\,i\in \{1,...,n-1\}
\end{equation*}%
and 
\begin{equation*}
e_{i}e_{j}=-e_{j}e_{i}=\beta _{ij}e_{k},\,\,\beta _{ij}\in \mathbb{R}%
,\,\,\,\beta _{ij}\neq 0,\,\,\,i\neq j,\,\,\,i,j\in \{\,\,1,...n-1\},
\end{equation*}%
$\beta _{ij}$ and $e_{k}$ being uniquely determined by $e_{i}$ and $e_{j}.$

Algebras $\mathbb{A}_{t}\left( \mathbb{R}\right) ,~$obtained by the
Cayley-Dickson process, are \textit{power-associative} (i.\thinspace e. the
subalgebra $<x>$ of $\mathbb{A}_{t}\left( \mathbb{R}\right) $, generated by
any element $x\in \mathbb{A}_{t}\left( \mathbb{R}\right) $, is associative), 
\textit{flexible }(i.\thinspace e. $x(yx)=(xy)x=xyx$, for all $x,y\in 
\mathbb{A}_{t}\left( \mathbb{R}\right) )$ and in general it is
nonassociative.

For $t=2$ and $\gamma _{1}=\gamma _{2}=-1,$ we obtain the Quaternion
division algebra, $\mathbb{Q}\left( \mathbb{R}\right) ,$ for $t=3$ and $%
\gamma _{1}=\gamma _{2}=\gamma _{3}=-1,$ we obtain the Octonion division
algebra, $\mathbb{O}\left( \mathbb{R}\right) ,$ and for $t=4$ and $\gamma
_{1}=\gamma _{2}=\gamma _{3}=\gamma _{4}=-1,$ we obtain the Sedenion
algebra, $\mathbb{S}\left( \mathbb{R}\right) $. Due to the Hurwitz's
Theorem, for $t\geq 4,$ all obtained algebras are not division algebras
(i.e. we can find $a,b\in \mathbb{A}_{t}\left( \mathbb{R}\right) ,$ $a\neq
0,b\neq 0,$ such that $ab=0$).

Let $B=\{1,e_{2},...,e_{2^{t}}\}$ be the basis in $\mathbb{A}_{t}\left( 
\mathbb{R}\right) ,$ where $1$ is the unity.$~$If $x=x_{1}+\overset{2^{t}}{%
\underset{i=2}{\sum }}x_{i}e_{i}\in \mathbb{A}_{t}\left( \mathbb{R}\right) ,$
then its \textit{conjugate} is the element $\overline{x}=x_{1}-\overset{2^{t}%
}{\underset{i=2}{\sum }}x_{i}e_{i}$ and the \textit{norm} of the element $x$
is $\mathbf{n}\left( x\right) =x\overline{x}=\overline{x}x=\overset{2^{t}}{%
\underset{i=1}{\sum }}x_{i}^{2}.$ The norm $\mathbf{n,}$ in general, is not
multiplicative, i.e. for $x,y\in \mathbb{A}_{t}\left( \mathbb{R}\right) ,$
we have $\mathbf{n}\left( xy\right) \neq \mathbf{n}\left( x\right) \mathbf{n}%
\left( y\right) .$ The \textit{real part} of the element $x$ is $x_{1}$ and
its \textit{vector part} is $\overset{2^{t}}{\underset{i=2}{\sum }}%
x_{i}e_{i}\in \mathbb{A}_{t}\left( \mathbb{R}\right) .\medskip $

The set $\mathbb{A}_{t}\left( \mathbb{Z}\right) =\{z\in \mathbb{A}_{t}\left( 
\mathbb{R}\right) ~|~\ z=x_{1}+\overset{2^{t}}{\underset{i=2}{\sum }}%
x_{i}e_{i},x_{i}\in \mathbb{Z},i\in \{2,3,...,2^{t}\}\}$ is called the 
\textit{integer elements} of the real algebra $\mathbb{A}_{t}\left( \mathbb{R%
}\right) $. This set has a ring structure. (see [Ma, Be, Ga; 09])

Let $w=\alpha (1+\overset{2^{t}}{\underset{i=2}{\sum }}e_{i})\in \mathbb{A}%
_{t}\left( \mathbb{R}\right) ,\alpha \in \mathbb{R},$ and let $\mathbb{V=\{}%
a+bw~|~~a,b\in \mathbb{Z}\mathbb{\}}$. We note that $\mathbf{t}\left(
x\right) =2\alpha ,$ $\mathbf{n}\left( x\right) =2^{t}\alpha ^{2}$ and $%
w^{2}-2\alpha w+2^{t}\alpha ^{2}=0.~$Since the algebra $\mathbb{A}_{t}\left( 
\mathbb{R}\right) $ is a power associative algebra, it results that $\mathbb{%
V}$ is an associative and a commutative ring.\medskip

\textbf{Remark 2.1.} For $x\in \mathbb{V},$\textit{\ } we know that the
following properties are equivalent:

i)\textit{\ }$x$ is an invertible element in the algebra\textit{\ }$\mathbb{V%
}.$

ii) $\mathbf{n}\left( x\right) =1.$

iii) $x\in \{\pm 1\}.\medskip $

An element $x\in \mathbb{V}$ is a \textit{prime} element in $\mathbb{V}$ if $%
x$ is not an invertible element in $\mathbb{V}$ and if $x=ab,$ it results
that $a$ or $b$ is an invertible element in $\mathbb{V}.\medskip $

\textbf{Proposition 2.2.} i) \textit{For} $x,y\in \mathbb{V},$ \textit{we
have} $\mathbf{n}\left( xy\right) =\mathbf{n}\left( x\right) \mathbf{n}%
\left( y\right) .$

ii) \textit{The ring}$~\mathbb{V}$ \textit{is a division ring}.\medskip

\textbf{Proof.} i) Denoting with $q=2^{t}-1,$ let $x=a+bw$ and $y=c+dw.$ We
obtain\newline
$\mathbf{n}\left( x\right) \mathbf{n}\left( y\right) =\left[ \left(
a+b\alpha \right) ^{2}+b^{2}\alpha ^{2}q\right] $ $\left[ \left( c+d\alpha
\right) ^{2}+d^{2}\alpha ^{2}q\right] =\newline
=\left( 2ab\alpha +a^{2}+b^{2}\alpha ^{2}+b^{2}q\alpha ^{2}\bigskip \right)
\left( 2cd\alpha +c^{2}+d^{2}\alpha ^{2}+d^{2}q\alpha ^{2}\right)
=\allowbreak 2abc^{2}\alpha \mathbf{+}2a^{2}cd\alpha +4abcd\alpha
^{2}+a^{2}c^{2}+2ab\allowbreak d^{2}\alpha ^{3}+2b^{2}cd\alpha
^{3}+2abd^{2}q\alpha ^{3}+2b^{2}cdq\alpha ^{3}+\allowbreak a^{2}d^{2}\alpha
^{2}+b^{2}c^{2}\alpha ^{2}+b^{2}d^{2}\alpha ^{4}+a^{2}d^{2}q\alpha
^{2}+b^{2}\allowbreak c^{2}q\alpha ^{2}+2b^{2}d^{2}q\alpha
^{4}+b^{2}d^{2}q^{2}\alpha ^{4}.$

Computing $\mathbf{n}\left( xy\right) ,$ we get

$\mathbf{n}\left( xy\right) =\left[ ac+\left( ad+bc\right) \alpha -\alpha
^{2}bd\left( q+1\right) +2\alpha ^{2}bd\right] ^{2}+q\alpha ^{2}\left[
ad+bc+2\alpha bd\right] ^{2}=\allowbreak 2abc^{2}\alpha +2a^{2}cd\alpha
+4abcd\alpha ^{2}+a^{2}c^{2}+2ab\allowbreak d^{2}\alpha ^{3}+2b^{2}cd\alpha
^{3}+2abd^{2}q\alpha ^{3}+2b^{2}cdq\alpha ^{3}+\allowbreak a^{2}d^{2}\alpha
^{2}+b^{2}c^{2}\alpha ^{2}+b^{2}d^{2}\alpha ^{4}+a^{2}d^{2}q\alpha
^{2}+b^{2}\allowbreak c^{2}q\alpha ^{2}+2b^{2}d^{2}q\alpha
^{4}+b^{2}d^{2}q^{2}\alpha ^{4}.$

$\allowbreak \allowbreak \allowbreak \allowbreak \allowbreak \allowbreak
\allowbreak \allowbreak \allowbreak $Therefore $\mathbf{n}\left( xy\right) =%
\mathbf{n}\left( x\right) \mathbf{n}\left( y\right) .$

ii) \ It results from i).\medskip

\textbf{Remark 2.3. }\ The above result is also true for all elements from
the set $\mathbb{V}^{\prime }\mathbb{=\{}a+bw~|~~a,b\in \mathbb{R\}}.$

In the following, we will consider $\alpha =\frac{1}{2^{r}},r\geq t-1,t\geq
2.\medskip $

\textbf{Proposition 2.4. } \textit{If} $x,y\in \mathbb{V},$\textit{\ }$y\neq
0,$\textit{\ with }$t\geq 2,~$\textit{then there are} $z,v\in \mathbb{V}$ 
\textit{such that} $x=zy+v$ and $\mathbf{n}\left( v\right) <\mathbf{n}\left(
y\right) .\medskip $

\textbf{Proof. }Since $y\neq 0\,,$ we have that $y$ is an invertible element
in $\mathbb{A}_{t}\left( \mathbb{R}\right) $, therefore $\frac{x}{y}%
=a+bw,a,b $ $\in \mathbb{R}.$ Let $m,n\in \mathbb{Z}$ such that $\left\vert
a-m\right\vert \leq \frac{1}{2}$ and $\left\vert b-n\right\vert \leq \frac{1%
}{2}.$ For $z=m+nw\in \mathbb{V}$ and $v=y\left[ \left( a-m\right) +\left(
b-n\right) w\right] ,$ it results that $\frac{x}{y}=z+\frac{v}{y},$
therefore $x=zy+v$ and $v=x-zy.$ From here, we have that $v\in \mathbb{V}$.
We obtain that\newline
$\mathbf{n}\left( v\right) =\mathbf{n}\left( y\right) \mathbf{n}\left(
\left( a-m\right) +\left( b-n\right) w\right) =$\newline
$=\mathbf{n}\left( y\right) \left[ \left[ \left( a-m\right) +\frac{1}{2^{r}}%
\left( b-n\right) \right] ^{2}+\frac{2^{r+1}-1}{2^{2r}}\left( b-n\right) ^{2}%
\right] \leq (\frac{(2^{r}+1)^{2}}{2^{2r+2}}+\frac{2^{r+1}-1}{2^{2r+2}})%
\mathbf{n}\left( y\right) =$\newline
$=\frac{2^{2r}+2^{r+2}}{2^{2r+2}}\mathbf{n}\left( y\right) =\frac{2^{r}+2^{2}%
}{2^{r+2}}\mathbf{n}\left( y\right) <\mathbf{n}\left( y\right) .\Box
\medskip $

\textbf{Definition 2.6. }With the above notations, let $\pi =x+yw$ be a
prime integer in $\mathbb{V}$ and $v_{1},v_{2}$ be two elements in $\mathbb{V%
}.$ If there is $v\in \mathbb{V}$~such that $v_{1}-v_{2}=v\pi ,$ then $%
v_{1},v_{2}$ are called \textit{congruent modulo} $\pi $ and it is denoted
by \ $v_{1}\equiv v_{2}$ \textit{mod} $\pi .\medskip $

\textbf{Proposition 2.7.} \textit{The above relation is an equivalence
relation on }$\mathbb{V}$\textit{. The set of equivalence classes is denoted
by} $\mathbb{V}_{\pi }$ \textit{and is called\ the residue classes of } $%
\mathbb{V}$\textit{\ modulo} $\pi .\smallskip $

\textbf{Proof.} Denoting the elements from $\mathbb{V}_{\pi }$ in bold$,$ if 
$v_{1}\equiv v_{2}$ \textit{mod} $\pi $ and $v_{2}\equiv v_{3}$ \textit{mod} 
$\pi ,$ then there are $v,v^{\prime }\in \mathbb{V}$ such that $%
v_{1}-v_{2}=v\pi $ and $v_{2}-v_{3}=v^{\prime }\pi .$ It results that $%
v_{1}-v_{3}=(v+v^{\prime })\pi ,~$therefore the transitivity holds. $\Box
\medskip $

\textbf{Proposition 2.8.} \textit{For each} $x,y\in \mathbb{V},$ \textit{%
there is} $\delta =\left( x,y\right) ,$ \textit{the greatest common divisor
of} $x$ \textit{and} $y$. \textit{We also have that there are} $\gamma $ 
\textit{and} $\tau \in \mathbb{V},$ \textit{such that} $\delta =\gamma
x+\tau y.$(the B\'{e}zout's Theorem).\smallskip

\textbf{Proof.} We denote by $J=\{\gamma x+\tau y\mid \gamma ,\tau \in 
\mathbb{V}\}.$We remark that if $z=\gamma ^{\prime }x+\tau ^{\prime }y\in J$
and $w\in \mathbb{V},$ we have $wz=(w\gamma ^{\prime })x+(w\tau ^{\prime
})y\in J.$ We consider $\delta _{1}=\gamma _{1}x+\tau _{1}y\in J,$ such that 
$\delta _{1}$ has the norm $\mathbf{n}\left( \delta _{1}\right) $ minimum in 
$J.$ We will prove that $\delta =\delta _{1}.$From Proposition 2.4, we have
that $x=q_{1}\delta _{1}+r_{1},$ with $\mathbf{n}\left( r_{1}\right) <%
\mathbf{n}\left( \delta _{1}\right) ,q_{1},r_{1}\in \mathbb{V}$ and $%
r_{1}=x-q_{1}\delta _{1}\in J.$ Since $\mathbf{n}\left( r_{1}\right) <%
\mathbf{n}\left( \delta _{1}\right) $ and $\delta _{1}\in J$ ~has minimum
norm in $J,$ it results that $r_{1}=0,$ therefore \thinspace $\delta
_{1}\mid x.\,$\ In the same way, we will prove that $\delta _{1}\mid y.$
Since $\delta _{1}=\gamma _{1}x+\tau _{1}y,$ it results that each common
divisor for $x$ and $y$ is a divisor for $\delta _{1},$ therefore $\delta
\mid \delta _{1}$ and finally $\delta =\delta _{1}.\Box \medskip $

The above proposition generalized to elements in $\mathbb{V}$ \ Proposition
2.1.4. from [Da, Sa, Va;03].\medskip

\textbf{Proposition 2.9.} $\mathbb{V}_{\pi }$ \textit{is a field } \textit{%
isomorphic to} $\mathbb{Z}/p\mathbb{Z},~p=\mathbf{n}(\pi )\medskip ,p$ 
\textit{a prime number.\medskip }

\textbf{Proof.}

For $\mathbf{v}_{1},\mathbf{v}_{2}\in \mathbb{V}_{\pi },~$we define $\mathbf{%
v}_{1}+\mathbf{v}_{2}=\left( v_{1}+v_{2}\right) $\textit{mod} $\pi $ and $%
\mathbf{v}_{1}\cdot \mathbf{v}_{2}=\left( v_{1}v_{2}\right) $\textit{mod} $%
\pi .$ These multiplications are well defined. Indeed, if $v_{1}\equiv
v_{1}^{\prime }$ \textit{mod} $\pi $ and $v_{2}\equiv v_{2}^{\prime }$ 
\textit{mod} $\pi ,$ it results that $v_{1}-v_{1}^{\prime }=u\pi
,v_{2}-v_{2}^{\prime }=u^{\prime }\pi ,u,u^{\prime }\in $ $\mathbb{V},$
therefore $\left( v_{1}+v_{2}\right) -\left( v_{1}^{\prime }+v_{2}^{\prime
}\right) =\left( u+u^{\prime }\right) \pi .$ Since $v_{1}=v_{1}^{\prime
}+u\pi ,$ $v_{2}=v_{2}^{\prime }+u^{\prime }\pi ,~$it results that $%
v_{1}v_{2}=v_{1}^{\prime }v_{2}^{\prime }+M_{\pi },$ with $M_{\pi }$ a
multiple of $\pi .$

Denoting in bold the equivalence classes from~ $\mathbb{Z}_{p},$ let $f$ \
be the map 
\begin{equation}
f:\mathbb{Z}_{p}\rightarrow \mathbb{V}_{\pi },~f\left( \mathbf{m}\right)
=\left( m+\pi \right) ~\text{\textit{mod }}\pi ,\text{ where }m\in \mathbf{m}%
.  \tag{2.1}
\end{equation}%
\newline
Map $f$ is well defined. Indeed, if $m\equiv m^{\prime }$ \textit{mod} $p$
we have $\left( m+\pi \right) -\left( m^{\prime }+\pi \right) =m-m^{\prime
}=pq=\pi \overline{\pi }q,q\in \mathbb{Z},$ therefore $\left( m+\pi \right)
\equiv \left( m^{\prime }+\pi \right) $ \textit{mod} $\pi .$ \newline

From Proposition 2.8, we have $1=v_{1}\pi +v_{2}\overline{\pi }.$ If $%
f\left( \mathbf{m}\right) =v,v=\left( m+\pi \right) ~$\textit{mod }$\pi \in 
\mathbb{V}_{\pi },$ we define $f^{-1}\left( v\right) =m\left( v_{1}\pi
\right) +m\left( v_{2}\overline{\pi }\right) =m.$

Map $f$ is a ring morphism. Indeed, $~f\left( \mathbf{m}\right) +f\left( 
\mathbf{m}^{\prime }\right) =\left( m+\pi \right) $\textit{mod} $\pi +\left(
m^{\prime }+\pi \right) $\textit{mod} $\pi =\left( m+m^{\prime }+\pi \right) 
$\textit{mod} $\pi =f\left( \mathbf{m}+\mathbf{m}^{\prime }\right) $ and%
\newline
$f\left( \mathbf{m}\right) f\left( \mathbf{m}^{\prime }\right) =\left( m+\pi
\right) \left( m^{\prime }+\pi \right) $\textit{mod}$\pi =$\newline
\textit{=}$\left( mm^{\prime }+\left( m+m^{\prime }\right) \pi +\pi
^{2}\right) $\textit{mod} $\pi =\left( mm^{\prime }+\pi \right) $\textit{mod}
$\pi .~$We obtain that $\mathbb{V}_{\pi }$ is isomorphic to $\mathbb{Z}%
_{p}.\Box \medskip $

Let $x=a+bw\in \mathbb{V},$ therefore we have $\mathbf{n}\left( x\right)
=\left( a+b\alpha \right) ^{2}+q(b\alpha )^{2}.$ For $q=2^{t}-1$ and for
certain values of $t,$ we know the form of some prime numbers, as we can see
in the proposition below.\medskip

\textbf{Proposition 2.9.} ([Co; 89], [Sa; 14]) \ 

\textit{Let} $p\in \mathbb{N}$ \textit{be a prime number. }

1) \textit{There are integers} $a,b$ \textit{\ such that} $p=a^{2}+3b^{2}$ 
\textit{\ if and only if} $p\equiv 1(mod$ $3)$ or $p=3.$

2) \textit{There are integers} $a,b$ \textit{\ such that} $p=a^{2}+7b^{2}$ 
\textit{\ if and only if} $p\equiv 1,2,4(mod$ $7)$ or $p=7.$

3) \textit{There are integers} $a,b$ \textit{\ such that} $p=a^{2}+15b^{2}$ 
\textit{\ if and only if} $p\equiv 1,19,31,49(mod~60).\Box \medskip $

\textbf{The label} \textbf{Algorithm for }$\mathbb{A}_{t}\left( \mathbb{R}%
\right) $\textbf{.}

1. We will fix $t,$ $\alpha $ and therefore $w.$

2. We consider $\pi $ $\in \mathbb{V}$ a prime element, $\pi =a+bw,a,b\in 
\mathbb{Z},$ such that $\mathbf{n}\left( \pi \right) =p=\left( a+b\alpha
\right) ^{2}+q(b\alpha )^{2},~$with $p$ a prime positive number$.$

3. Let $s\in \mathbb{Z}$ be the only solution to the equation $a+bx$ $=0$ 
\textit{mod} $p,~x\in \{0,1,2,...,p-1\}.$

4. Let $r=\left[ \frac{p-1}{2}\right] \in \mathbb{N},~$where $\left[ ~~%
\right] $ denotes the integer part.

5. Let $k\in \mathbb{Z\,\ }$and $\mathbf{k\in }\mathbb{Z}_{p}$ be its
equivalence class modulo $p.$

6. For all integers $\ \sigma ,\tau \in \mathbb{\{}-r-1,...,r\mathbb{\}},$
let $c=(s\tau +\sigma )$ \textit{mod} $p$ and $d=(\sigma +\tau \alpha
)^{2}+q(\tau \alpha )^{2}.$

6. If $d<p$ and $c=k,$ then we find the pairs $\left( \sigma ,\tau \right) $
such that $\mathbf{k}$ is the label of the element $\sigma +\tau w\in 
\mathbb{V}_{\pi },$ that means $\sigma +\tau s=k~$\textit{mod~}$p~$and$~%
\mathbf{n}\left( \sigma +\tau w\right) ~$is~minimum$.$ If we find more than
two pairs satisfying the last condition, then we will choose that pair with
the property that $\left\vert \sigma \right\vert +\left\vert \tau
\right\vert \leq \left\vert a\right\vert +\left\vert b\right\vert .$ If
there exist more than two pairs satisfying the last inequality, then we will
choose one of them randomly.\medskip 
\begin{equation*}
\end{equation*}

\bigskip

\textbf{3.} \textbf{Codes over } $\mathbb{V}_{\pi }$

\begin{equation*}
\end{equation*}

In the following, we will recall some definitions, which will be used in
this section.

We consider the ring of Gaussian integers, $\mathbb{Z}[i]=\{a+bi~\mid $ $%
a,b\in \mathbb{Z},i^{2}=-1\}$. We know that a prime integer $p$ of the form $%
p\equiv 1$ \textit{mod} $4$ can be written of the form \thinspace $p=\pi 
\overline{\pi },$ where $\pi ,\overline{\pi }\in \mathbb{Z}[i]$ and $%
\overline{\pi }$ is the conjugate of \ $\pi .$ Let $(\mathbb{Z}[i])_{\pi }$
be the set of the residue classes modulo $\pi .$ A block code $C$ of length $%
n$ over $(\mathbb{Z}[i])_{\pi }$ is defined to be a set of codewords $%
c=\left( c_{1},...,c_{n}\right) ,$ where $c_{i}\in (\mathbb{Z}[i])_{\pi },$ $%
i\in \{1,2,...,n\}.$ For $\alpha ^{\prime },\beta ^{\prime },\gamma ^{\prime
}\in (\mathbb{Z}[i])_{\pi },$ with $\gamma ^{\prime }=\alpha ^{\prime
}-\beta ^{\prime }$ mod $\pi ,$ the \textit{Mannheim weight} of $\gamma
^{\prime },$ denoted by $w_{M}\left( \gamma ^{\prime }\right) ,$ is defined
as 
\begin{equation*}
w_{M}\left( \gamma ^{\prime }\right) =\left\vert Re\text{(}\gamma ^{\prime }%
\text{)}\right\vert +\left\vert Im\text{(}\gamma ^{\prime }\text{)}%
\right\vert ,
\end{equation*}%
where $Re$($\gamma ^{\prime }$) represents the real part of the element $%
\gamma ^{\prime }$ and $Im$($\gamma ^{\prime }$) represents the imaginary
part of the element $\gamma ^{\prime }.$ Using the Mannheim weight, we can
define the \textit{Mannheim distance }between $\alpha ^{\prime }$ and $\beta
^{\prime },$ denoted by $d_{M},$ as follows 
\begin{equation*}
d_{M}\left( \alpha ^{\prime },\beta ^{\prime }\right) =w_{M}\left( \gamma
^{\prime }\right) .
\end{equation*}%
\qquad \qquad

For other details, the readers are referred to [Hu; 94] .

Using ideas from the above definition and generalizing the Hurwitz weight
from [Gu; 13] and Cayley-Dickson weight for the octonions,\ from [Fl; 15],
in the same manner, we define the \textit{generalized} \textit{%
Cayley-Dickson weight}, for algebras obtained by the Cayley-Dickson process,
denoted $d_{G}.$ We will fix $t,$ $\alpha $, $w,$ therefore we will consider
the elements in the algebra $\mathbb{A}_{t}\left( \mathbb{R}\right) .$ Let $%
\pi $ be a prime in $\mathbb{V},$ $\mathbb{\pi =}a+bw.$ Let $x\in \mathbb{V}%
, $ $x=a_{0}+b_{0}w.$ The \textit{generalized} \textit{Cayley-Dickson weight}
of $x$ is defined as $w_{G}\left( x\right) =\left\vert a_{0}\right\vert
+\left\vert b_{0}\right\vert ,~$where $x=a_{0}+b_{0}w$ \textit{mod}$~\pi ,$
with $\left\vert a_{0}\right\vert +\left\vert b_{0}\right\vert $ minimum. 
\newline

The \textit{generalized Cayley-Dickson distance }between $x,y\in \mathbb{V}%
_{\pi }$ is defined as 
\begin{equation*}
d_{G}\left( x,y\right) =w_{G}\left( x-y\right)
\end{equation*}%
and we will prove that $d_{G}$ is a metric. Indeed, for $x,y,z$ $\in $ $%
\mathbb{V}_{\pi },$ we have $d_{G}\left( x,y\right) =w_{G}\left( \alpha
_{1}\right) =\left\vert a_{1}\right\vert +\left\vert b_{1}\right\vert ,$
~where $\alpha _{1}=x-y=a_{1}+b_{1}w$ \textit{mod} $\pi $ is an element in $%
\mathbb{V}_{\pi }$ and $\left\vert a_{1}\right\vert +\left\vert
b_{1}\right\vert $ is minimum.\newline
$d_{G}\left( y,z\right) =w_{G}\left( \alpha _{2}\right) =\left\vert
a_{2}\right\vert +\left\vert b_{2}\right\vert ,~$where $\alpha
_{2}=y-z=a_{2}+b_{2}$ w\textit{mod} $\pi $ is an element in $\mathbb{V}_{\pi
}$ and $\left\vert a_{2}\right\vert +\left\vert b_{2}\right\vert $ is
minimum.\newline
$d_{G}\left( x,z\right) =w_{G}\left( \alpha _{3}\right) =\left\vert
a_{3}\right\vert +\left\vert b_{3}\right\vert ,~$where $\alpha
_{3}=x-z=a_{3}+b_{3}w$ \textit{mod} $\pi $ is an element in $\mathbb{V}_{\pi
}$ and $\left\vert a_{3}\right\vert +\left\vert b_{3}\right\vert $ is
minimum.\newline
We obtain $x-z=\alpha _{1}+\alpha _{2}$ \textit{mod} $\pi $ and it results
that $w_{G}\left( \alpha _{1}+\alpha _{2}\right) \geq w_{G}\left( \alpha
_{3}\right) ,$ since $w_{G}\left( \alpha _{3}\right) =\left\vert
a_{3}\right\vert +\left\vert b_{3}\right\vert $ is minimum, therefore $%
d_{G}\left( x,y\right) +d_{G}\left( y,z\right) \geq d_{G}\left( x,z\right) $%
.\medskip \medskip

In the following, we assume that $\pi $ is a prime in $\mathbb{V}$ with $%
\mathbf{n}\left( \pi \right) =p$ a prime number of the form $\mathbf{n}%
\left( \pi \right) =Mn+1,\ M,n\in \mathbb{Z},n\geq 0,\ $such that there are $%
\beta $ a primitive element (of order $p-1)$ in $\mathbb{V}_{\pi },$ with
the properties $\beta ^{\frac{p-1}{M}}=w$ or $\beta ^{\frac{p-1}{M}}=-w.$ We
will consider codes of length $n=\frac{p-1}{M}.$

The below definitions and Theorems adapted and generalized to all algebras
obtained by the Cayley-Dickson process some definitions from [Gu; 13], [Ne,
In, Fa, Pa; 01], [Fl; 15] and Theorems 7,8,9,10,11,13,14,15 from [Ne, In,
Fa, Pa; 01], Theorems 4,5,6,7 from [Gu; 13] and Theorems 2.3, 2.5, 2.7, 2.9
from [Fl; 15].

Let $C$ be a code defined by the parity-check matrix $H,$%
\begin{equation}
H=\left( 
\begin{array}{ccccc}
1 & \beta & \beta ^{2} & ... & \beta ^{n-1} \\ 
1 & \beta ^{M+1} & \beta ^{2(M+1)} & ... & \beta ^{(n-1)(M+1)} \\ 
... & ... & ... & ... & ... \\ 
1 & \beta ^{Mk+1} & \beta ^{2\left( Mk+1\right) } & ... & \beta ^{\left(
n-1\right) (Mk+1)}%
\end{array}%
\right) ,  \tag{3.1}
\end{equation}%
with $k<n.$ We know that $c$ is a codeword in $C$ if and only if $Hc^{t}=0.$
If \ we consider the associate code polynomial $c\left( x\right) =\underset{%
i=0}{\overset{n-1}{\sum }}c_{i}x^{i},$ we have that $c\left( \beta
^{Ml+1}\right) =0,l\in \{0,1,...,k\}.$ For the polynomial $g\left( x\right)
=\left( x-\beta \right) \left( x-\beta ^{M+1}\right) ...\left( x-\beta
^{(Mk+1)}\right) ,$ since the elements $\beta ,\beta ^{M+1},...,\beta
^{Mk+1} $ are distinct, from [10], Lemma 8.1.6, we obtain that $c\left(
x\right) $ is divisible by $g\left( x\right) ,$ the generator polynomial of
the code $C. $ Since $g\left( x\right) ~/~($ $x^{n}\pm w)$, it results that $%
C$ is a principal ideal in the ring $\mathbb{V}_{\pi }~/$ $(x^{n}\pm w).$

If we suppose that a codeword polynomial $c\left( x\right) ~$is sent over a
channel and the error pattern $e\left( x\right) $ occurs, it results that
the received polynomial is $r\left( x\right) =c\left( x\right) +e\left(
x\right) .$ The vector corresponding to the polynomial $r\left( x\right)
=c\left( x\right) +e\left( x\right) $ is $r=c+e$ and \textit{the syndrome}
of $r$ is $S=Hr^{t},$ where $H$ is the above parity-check matrix.$\medskip $

\textbf{Theorem 3.1.} \textit{Let } $C$ \textit{be the code defined on }$%
\mathbb{V}_{\pi }$\textit{\ by the parity check matrix }%
\begin{equation}
H=\left( 
\begin{array}{ccccc}
1 & \beta & \beta ^{2} & ... & \beta ^{n-1}%
\end{array}%
\right) .  \tag{3.2}
\end{equation}%
\textit{It results that, the code} $C$ \textit{is able to correct all error
patterns of the form} $e\left( x\right) =e_{i}x^{i},$ \textit{with} $0\leq
w_{C}\left( e_{i}\right) \leq 1.$ $\medskip $

\textbf{Proof.} Let $r\left( x\right) =c\left( x\right) +e\left( x\right) $
be the received polynomial, with $c(x)$ the codeword polynomial and $%
e(x)=e_{i}x^{i}$ denoting the error polynomial with $0\leq w_{C}\left(
e_{i}\right) \leq 1.$ Since $\beta ^{n}=w$ or $\beta ^{n}=-w,$ it results
that $e_{i}=\beta ^{nl}.$ We have the syndrome $S=\beta ^{i+nl}=\beta ^{L},$
with $i,L\in \mathbb{Z},0\leq i,L\leq n-1.$ If we reduce $L$ modulo $n,$ we
obtain $i,$ the location of the error, and from here, $l=\frac{L-i}{n}$ and $%
\beta ^{nl},~$the value of the error.$~\Box \medskip $

\textbf{Theorem 3.2.} \textit{Let }$\ C$ \textit{be a code defined by the
parity-check matrix} \newline
\begin{equation}
H=\left( 
\begin{array}{ccccc}
1 & \beta & \beta ^{2} & ... & \beta ^{n-1} \\ 
1 & \beta ^{M+1} & \beta ^{2(M+1)} & ... & \beta ^{(n-1)(M+1)}%
\end{array}%
\right) .  \tag{3.3}
\end{equation}%
\textit{Then} $C$ \textit{can correct error patterns of the form} $e\left(
x\right) =e_{i}x^{i},$ $0\leq i\leq n-1,$ \textit{with}$~e_{i}\in \mathbb{V}%
_{\pi }.\medskip $

\textbf{Proof.} We consider the received polynomial, $r\left( x\right)
=c\left( x\right) +e\left( x\right) $ with $c(x)$ the codeword polynomial
and $e(x)=e_{i}x^{i}$ the error polynomial with $e_{i}\in \mathbb{V}_{\pi }.$
It results that the corresponding vector of the polynomial $r(x)$ is $r=c+e.$
We will compute the syndrome $S$ of $r.$ We have $e_{i}=\beta ^{j},0\leq
j\leq Mn-1.$ Therefore the syndrome is 
\begin{equation*}
S\text{=}Hr^{t}\text{=}\left( 
\begin{array}{c}
s_{1}=\beta ^{i+j}=\beta ^{M_{1}} \\ 
s_{M+1}=\beta ^{(M+1)i+j}=\beta ^{M_{2}}%
\end{array}%
\right) .
\end{equation*}%
We obtain $\beta ^{i+j-M_{1}}=1,$ with $i+j=M_{1}$ \textit{mod}$(p-1)$ and $%
\beta ^{(M+1)i+j-M_{2}}=1,$ with $(M+1)i+j=M_{2}$ \textit{mod}$(p-1).$ We
get $Mi=(M_{2}-M_{1})$ \textit{mod}$(p-1),$ if there is, then the solution
of the system is $i=\frac{M_{2}-M_{1}}{M}$ \textit{mod~}$n$ and $j=(M_{1}-i)$
\textit{mod}$(p-1).$ From here, we can find the location and the value of
the error.$\Box \medskip $

\textbf{Theorem 3.3.} \textit{Let }$\ C$ \textit{be a code defined by the
parity-check matrix} \newline
\begin{equation}
H=\left( 
\begin{array}{ccccc}
1 & \beta & \beta ^{2} & ... & \beta ^{n-1} \\ 
1 & \beta ^{M+1} & \beta ^{2(M+1)} & ... & \beta ^{(n-1)(M+1)} \\ 
1 & \beta ^{2M+1} & \beta ^{2(2M+1)} & ... & \beta ^{(n-1)(2M+1)}%
\end{array}%
\right) .  \tag{3.4}
\end{equation}%
\textit{Then} $C$ \textit{can find the location and can correct error
patterns of the form} $e\left( x\right) =e_{i}x^{i},$ $0\leq i\leq n-1,$ 
\textit{with}$~e_{i}\in \mathbb{V}_{\pi },$ \textit{or can only correct
error patterns of the above mentioned form}$.\medskip $

\textbf{Proof.} Using notations from the above Theorem, we have $e_{i}=\beta
^{j},0\leq j\leq Mn-1.$ Therefore the syndrome is 
\begin{equation*}
S\text{=}Hr^{t}\text{=}\left( 
\begin{array}{c}
s_{1}=\beta ^{i+q}=\beta ^{M_{1}} \\ 
s_{M+1}=\beta ^{\left( M+1\right) i+j}=\beta ^{M_{2}} \\ 
s_{2M+1}=\beta ^{\left( 2M+1\right) i+j}=\beta ^{M_{3}}%
\end{array}%
\right) .
\end{equation*}

Since the matrix $\left( 
\begin{array}{ccc}
1 & \beta & \beta ^{2} \\ 
1 & \beta ^{M+1} & \beta ^{2(M+1)} \\ 
1 & \beta ^{2M+1} & \beta ^{2(2M+1)}%
\end{array}%
\right) $ has its determinant equal to $\beta ^{3}\beta ^{M}\left( \beta
^{2M}-1\right) ^{3}\neq 0,$ it results that the rank of the matrix $\left(
3.4\right) $ is $3,$ then this system always has a solution. We obtain $%
\beta ^{i+j-M_{1}}=1,$ with $i+q=M_{1}$ \textit{mod}$(p-1),$ $\beta ^{\left(
M+1\right) i+j-M_{2}}=1,$ with $\left( M+1\right) i+j=M_{2}$ \textit{mod}$%
(p-1),~\beta ^{\left( 2M+1\right) i+j-M_{3}}=1,~$with $\left( 2M+1\right)
i+j=M_{3}~$\textit{mod}$(p-1).$ We can find the location of the error if $%
Mi=(M_{2}-M_{1})$ \textit{mod}$(p-1)$ and $Mi=(M_{3}-M_{2})$ \textit{mod}$%
(p-1)$ or, equivalently, $i=\frac{M_{2}-M_{1}}{M}$ \textit{mod~}$n=\frac{%
M_{3}-M_{2}}{M}$ \textit{mod~}$n$ and the value of the error $e_{i}$ if 
\newline
$(M_{1}-i)$ \textit{mod}$(p-1)=(M_{2}-\left( M+1\right) i)$ \textit{mod}$%
(p-1)=(M_{3}-\left( 2M+1\right) i)$ \textit{mod}$(p-1)(=j).$ $\Box \medskip $

\textbf{Theorem 3.9.} \textit{Let }$\ C$ \textit{be a code defined by the
parity-check matrix} \newline
\begin{equation}
H=\left( 
\begin{array}{ccccc}
1 & \beta & \beta ^{2} & ... & \beta ^{n-1} \\ 
1 & \beta ^{M+1} & \beta ^{2(M+1)} & ... & \beta ^{(n-1)(M+1)} \\ 
1 & \beta ^{2M+1} & \beta ^{2(2M+1)} & ... & \beta ^{(n-1)(2M+1)} \\ 
1 & \beta ^{3M+1} & \beta ^{2\left( 3M+1\right) } & ... & \beta ^{\left(
n-1\right) \left( 3M+1\right) }%
\end{array}%
\right) .  \tag{3.5}
\end{equation}%
\textit{Then} $C$ \textit{can correct error patterns of the form} $e\left(
x\right) =e_{i}x^{i}+e_{j}x^{j},$ $0\leq i,j\leq n-1,$ \textit{with}$%
~e_{i},e_{j}\in \mathbb{V}_{\pi }.\medskip $

\textbf{Proof. }\ We will prove this in the general case, when we have two
errors. We have $e_{i}=\beta ^{q\prime }\neq 0$ and $e_{j}=\beta ^{t^{\prime
}}\neq 0,q^{\prime },t^{\prime }\in \mathbb{Z}$. We obtain the syndrome:%
\begin{equation*}
S\text{=}Hr^{t}\text{=}\left( 
\begin{array}{c}
s_{1}=\alpha ^{i+q^{\prime }}+\alpha ^{j+t\prime } \\ 
s_{M+1}=\alpha ^{\left( M+1\right) i+q^{\prime }}+\alpha ^{\left( M+1\right)
j+t^{\prime }} \\ 
s_{2M+1}=\alpha ^{\left( 2M+1\right) i+q^{\prime }}+\alpha ^{(2M+1)j+t\prime
} \\ 
s_{3M+1}=\alpha ^{\left( 3M+1\right) i+q^{\prime }}+\alpha ^{\left(
3M+1\right) j+t^{\prime }}%
\end{array}%
\right) .
\end{equation*}%
We denote $\beta ^{i+q^{\prime }}=A$ and $\beta ^{j+t^{\prime }}=B$ and we
get

\begin{equation}
S\text{=}Hr^{t}\text{=}\left( 
\begin{array}{c}
s_{1}=A+B \\ 
s_{M+1}=\beta ^{Mi}A+\beta ^{Mj}B \\ 
s_{2M+1}=\beta ^{2Mi}A+\beta ^{2Mj}B \\ 
s_{3M+1}=\beta ^{3Mi}A+\beta ^{3Mj}B%
\end{array}%
\right) .  \tag{3.6}
\end{equation}

If the system $\left( 3.6\right) $ admits only one solution, then code $C$
can correct two errors. First, we will prove the following Lemma.\medskip

\textbf{Lemma.} \ \textit{With the above notations, if we have two errors,
we obtain} $\beta ^{Mi}\neq \beta ^{Mj},0\leq i,j\leq n-1$ \textit{and} $%
s_{1}s_{2M+1}\neq s_{M+1}^{2}.\medskip $

\textbf{Proof.} If \ $\beta ^{Mi}=\beta ^{Mj},~$then $\beta ^{M(i-j)}=1$ and 
$Mn~/~M(i-j),$ which is false. Supposing that $s_{1}s_{2M+1}-s_{M+1}^{2}=0,$
we have $s_{1}s_{2M+1}=s_{M+1}^{2}.$ $\ $For $x=$ $\beta ^{i+q^{\prime }},$
we obtain that $\beta ^{2Mi}s_{1}x+\beta ^{2Mj}s_{1}^{2}-\beta
^{2Mj}s_{1}x=\left( \beta ^{Mi}-\beta ^{Mj}\right) ^{2}x^{2}+\beta
^{2Mj}s_{1}^{2}+2\beta ^{Mj}\left( \beta ^{Mi}-\beta ^{Mj}\right) s_{1}x.$
It results $(\beta ^{Mi}-\beta ^{Mj})^{2}x^{2}+2\beta ^{Mi+Mj}s_{1}x-\beta
^{2Mi}s_{1}x-\beta ^{2Mj}s_{1}x=0.$ From here, $x=0$ or $x=\frac{-2\beta
^{Mi+Mj}s_{1}+\beta ^{2Mi}s_{1}+\beta ^{2Mj}s_{1}}{\left( \beta ^{Mi}-\beta
^{Mj}\right) ^{2}}=s_{1}.$ If we have $x=$ $\beta ^{i+q^{\prime }}=s_{1},$
this implies $\beta ^{j+t^{\prime }}=0,~$which is false.\medskip

We now return to the proof of the Theorem and we know that the following
conditions are fulfilled: $\beta ^{Mi}\neq \beta ^{Mj},0\leq i,j\leq n-1$
and $s_{1}s_{2M+1}\neq s_{M+1}^{2}.\medskip $ For $B=s_{1}-A,$ it results
that\newline
$A\left( \beta ^{Mi}-\beta ^{Mj}\right) =s_{M+1}-s_{1}\beta ^{Mj}$\newline
$A\left( \beta ^{2Mi}-\beta ^{2Mj}\right) =s_{2M+1}-s_{1}\beta ^{2Mj}$%
\newline
$A\left( \beta ^{3Mi}-\beta ^{3Mj}\right) =s_{3M+1}-s_{1}\beta ^{3Mj}$. 
\newline
We obtain $s_{2M+1}-s_{1}\beta ^{2Mj}=\left( s_{M+1}-s_{1}\beta ^{Mj}\right)
\left( \beta ^{Mi}+\beta ^{Mj}\right) $ and $s_{3M+1}-s_{1}\beta
^{3Mj}=\left( s_{M+1}-s_{1}\beta ^{Mj}\right) \left( \beta ^{2Mi}+\beta
^{Mi}\beta ^{Mj}+\beta ^{2Mj}\right) .$\newline
If we denote by $s_{M}=\beta ^{Mi}+\beta ^{Mj}$ and $p_{M}=\beta ^{Mi}\beta
^{Mj},$ we have

\begin{equation*}
s_{2M+1}-s_{M+1}s_{M}+p_{M}s_{1}=0
\end{equation*}%
and 
\begin{equation*}
\left( s_{M+1}-s_{1}\beta ^{Mj}\right) \left( s_{M}^{2}-p_{M}\right)
=s_{3M+1}-s_{1}\beta ^{3Mj}.
\end{equation*}

It results that%
\begin{equation*}
p_{M}=\frac{s_{M+1}s_{M}-s_{2M+1}}{s_{1}}
\end{equation*}%
and%
\begin{equation*}
s_{M}(s_{1}s_{2M+1}-s_{M+1}^{2})=s_{1}s_{3M+1}-s_{M+1}s_{2M+1}.
\end{equation*}

Therefore, we obtain 
\begin{equation*}
s_{M}=\frac{s_{1}s_{3M+1}-s_{M+1}s_{2M+1}}{s_{1}s_{2M+1}-s_{M+1}^{2}}
\end{equation*}
\begin{equation*}
p_{M}=\frac{s_{M+1}s_{3M+1}-s_{2M+1}^{2}}{s_{1}s_{2M+1}-s_{M+1}^{2}}.
\end{equation*}

From here, by solving the equation $x^{2}-s_{M}x+p_{M}=0,$ we find the
locations and the values of the errors. $\Box \medskip $\newline

\begin{equation*}
\end{equation*}

\textbf{4. Main algorithm and some examples}%
\medskip

\textbf{Definition 4.1. }\textit{The rate} of a block code is $R=\frac{k}{n}$%
, \ where $k$ is the dimension of the block code and $n~$is the length of
the codewords.\bigskip

\textbf{The Main Algorithm\medskip \medskip }

Let $p$ be a prime number.

1. We find $a,b,t\in \mathbb{N}$ such that we can write $p$ under the form%
\begin{equation}
p=a^{2}+\left( 2^{t}-1\right) b^{2}.  \tag{4.1.}
\end{equation}%
We remark that the values for $a,b,t$, \ if there exist, are not unique. Let 
$\{a_{l},b_{l},t_{l}\},l\in \{1,2,...,u\}$ all solutions of the equation $%
\left( 4.1\right) .$

2. Let $p=n_{j}M_{j}+1,~$with $n_{j},M_{j}$ not unique such that $%
n_{j}M_{j}=p-1,j\in \{1,2,...,v\}.$

3. For $l\in \{1,2,...,u\}$ and for $j\in \{1,2,...,v\},~$we find the
algebra $\mathbb{A}_{t_{l}}\left( \mathbb{R}\right) ,~$the element $w=\frac{1%
}{2^{r_{l}-1}}(1+\overset{2^{t_{l}}}{\underset{i=2}{\sum }}e_{i})\in \mathbb{%
A}_{t_{l}}\left( \mathbb{R}\right) ,r_{l}\geq t_{l}-1,\mathbb{V\subset A}%
_{t_{l}}\left( \mathbb{R}\right) ,$ the element$\ \pi \in \mathbb{V},$ such
that $\mathbf{n}\left( \pi \right) =p,$ we find $\mathbb{V}_{\pi }$ such
that $\mathbb{V}_{\pi }$ is isomorphic to $\mathbb{Z}_{p}$ and we find $%
\beta \in \mathbb{V}_{\pi }$ such that $\beta ^{n_{j}}=w$ or $\beta
^{n_{j}}=-w.$

If the elements $\{a_{l},b_{l},t_{l}\}$ don't exist, then the algorithm \
stops.

If we have at least a solution for the equation $\left( 4.1\right) $ but we
don't find for $j\in \{1,2,...,v\}$ the element $\beta \in \mathbb{V}_{\pi }$
such that $\beta ^{n_{j}}=w$ or $\beta ^{n_{j}}=-w,$ then the algorithm \
stops. If we have solutions in both cases, then we go to the Step 4.

4. For each solution $\{a_{l},b_{l},t_{l}\},l\in \{1,2,...,u\},$ let $%
\mathcal{J}\subseteq \{1,2,...,v\}.$ For each $j\in \mathcal{J},~$we have $%
n_{j}$ such that $\beta ^{n_{j}}=w$ or $\beta ^{n_{j}}=-w.$ We can change $w$
by increasing the value of $r_{l\text{ }},~$if it is necessary, but working
in the algebra $\mathbb{A}_{t_{l}}\left( \mathbb{R}\right) $. For each $%
n_{j} $ we compute $M_{j}$ and the rate of the obtained code, $R_{j}=\frac{%
k_{j}}{n_{j}}.$ Since we can suppose that the obtained codes have the same
dimension $k=k_{j},$ we will chose the indices $l\in \{1,2,...,u\}$, $j\in 
\mathcal{J},~$the pair $\{a_{l},b_{l},t_{l}\}$ and the number $n_{j}$ such
that the rate $R_{j}$ has the biggest value.\medskip

In the following, we will denote by Algorithm 1, the method described in
[Gu; 13] and by Algorithm 2, the method described in [Fl; 15].\medskip

\textbf{Remark 4.2.} In the papers [Gu; 13] and [Fl; 15] were developed some
algorithms which built binary block codes over subsets of integers in the
real quaternion division algebra and in the real octonion division algebra.
The above algorithm generalized these two algorithms to real algebras
obtained by the Cayley-Dickson process. Moreover, the Main Algorithm can be
generalized to almost all prime number, which in general the Algorithm 1 and
the Algorithm 2 don't make it. That means, in general, for a prime number $p$%
, we can get the algebra $\mathbb{A}_{t}\left( \mathbb{R}\right) ,$ the
element $w\in \mathbb{A}_{t}\left( \mathbb{R}\right) ,$ the subset $\mathbb{%
V\subset A}_{t_{l}}\left( \mathbb{R}\right) ,$ $\pi \in \mathbb{V},$ such
that $\mathbf{n}\left( \pi \right) =p,$ we can find $\mathbb{V}_{\pi }$ with 
$\mathbb{V}_{\pi }$ isomorphic to $\mathbb{Z}_{p},$ such that the obtained
binary block code can have the highest rate.

With the above algorithm, we have a much higher flexibility, similar to the
Lenstra algorithm for elliptic curves compared with $p-1$ Pollard algorithm.
It is well known that for a prime $p,$ Lenstra's algorithm replace the group 
$\mathbb{Z}_{p}^{\ast }$ with the group of the rational points of an
elliptic curve $\mathcal{C}_{1}$ over $\mathbb{Z}_{p}$ and if \ this
algorithm failed, the curve will be replaced with another curve $\mathcal{C}%
_{2}$ over $\mathbb{Z}_{p}$ and we can retake the algorithm (see [Si, Ta;
92]).

In the case of Main Algorithm, the algebra $\mathbb{A}_{t}\left( \mathbb{R}%
\right) $ and $w$ offer this kind of flexibility since, for the same prime $%
p,$ these can be changed and the algorithm can be retake, with better
chances of success.\medskip

We will explain this in the following examples.\medskip\ 

\textbf{Example 4.3.} Let $p=29.$ We have $a=1,b=2$ and $\ t=3,$ therefore $%
p=1+7\cdot 4$ with unique decomposition. It results that we can use the real
Octonion algebra. If we apply Algorithm 2, we have $\ w=\frac{1}{2}\left( 1+%
\overset{8}{\underset{i=2}{\sum }}e_{i}\right) ,$ $\pi
=-1+4w,p=29,n=4,s=22,\beta =1-w,\beta ^{4}=-w$ \textit{mod} $\pi ,$
therefore we can define codes$.$

If we apply the Main Algorithm for $w=\frac{1}{4}\left( 1+\overset{8}{%
\underset{i=2}{\sum }}e_{i}\right) ,$ we have $\pi =-1+8w,n=4,s=11$ which is
the label for the element $w\in \mathbb{V}_{\pi }.$ We remark that we can't
find $\beta \in \mathbb{V}_{\pi }$ such that $\beta ^{4}=w$, as we can see
from the MAPLE's procedures below.\medskip
\begin{verbatim}
for i from -15 to 14 do for j from -15 to 14 do 
c := (11*j+i)mod 29; d := ((7/4)*j)^2+(i+(1/4)*j)^2; 
if d < 29 and c = 11 then print(i, j);fi;od;od;
                                    0, 1
                                    4, -2
 
A := 8^{-1} mod 29; for a to 29 do 
b := a^4 mod 29; if b = 11 then print(a);fi;od;
                                     11
 
\end{verbatim}

\bigskip But, if we increase $\alpha $ we still work on octonions and we
take $w=\frac{1}{32}\left( 1+\overset{8}{\underset{i=2}{\sum }}e_{i}\right)
, $ with the label $s=24.$ We obtain $\beta =-1-w$ with the label $4~$such
that $\beta ^{4}=w,$therefore we can define codes. In this situation, both
algorithms can be applied with success.\medskip\ 

\textbf{Example 4.4. }Let $p=71=64+7\cdot 1,\ $with unique decomposition.
Therefore $a=8,b=1,t=3.$ Then we work on real Octonion algebra. If we apply
the Algorithm 2, we have $w=\frac{1}{2}\left( 1+\overset{8}{\underset{i=2}{%
\sum }}e_{i}\right) ,$ $\pi =7+2w,p=71,n=10,s=32,\beta =2-2w,\beta ^{10}=w$ 
\textit{mod} $\pi .($see [Fl; 15])

If we apply the Main Algorithm and first we take $w=\frac{1}{4}\left( 1+%
\overset{8}{\underset{i=2}{\sum }}e_{i}\right) ,$ we have $\pi
=7+4w,p=71,n=10,s=16$ which is the label for the element $w\in \mathbb{V}%
_{\pi }.$We remark that we can't find $\beta \in \mathbb{V}_{\pi }$ such
that $\beta ^{10}=w$ (even if we increase the value of $r,$ as in Example
4.3), as we can see in the procedure below.\medskip
\begin{verbatim}
A := -7*4^{-1}mod71; for a from 1 to 71 do b := a^{10} mod 71; 
if b = 16 then print(a);fi;od:
                                     16
 
\end{verbatim}

Therefore, the Algorithm 2 is better than the Main Algorithm.\medskip

\textbf{Example 4.5}. For $p=31=6\cdot 5+1,$ we have $p=4+3\cdot
9=16+15\cdot 1,$ therefore $t\in \{4,16\}$ and we can use the real
Quaternion algebra or the real Sedenion algebra. If we apply the Main
Algorithm for sedenions, we have $w=\frac{1}{8}\left( 1+\overset{16}{%
\underset{i=2}{\sum }}e_{i}\right) .$ We get $\pi =3+8w,p=31,s=19$ and we
remark that we can't use it for the sedenions since we can't find $\beta \in 
\mathbb{V}_{\pi }$ such that $\beta ^{5}=w.$ Therefore, we will use the Main
Algorithm only for Quaternion algebra, which can be applied in this
case.\medskip

\textbf{Example 4.6. }\ Let $p=61.$ We have that $p=4\cdot 3\cdot
5+1=1+60=1+15\cdot 4=49+3\cdot 4,$ therefore $t\in \{4,16\}$ and we can use
the real Quaternion algebra or the real Sedenion algebra.

If \ \ we take $p$ under the form $p=61=7^{2}+3\cdot 2^{2},$ \ we use the
real Quaternion algebra$.$ For $w=\frac{1}{2}\left( 1+\overset{4}{\underset{%
i=2}{\sum }}e_{i}\right) ,$ we get $\pi =5+4w,$ the label for $w$ is $%
s=14,n=10(p=6\cdot 10+1)$ and we have $\beta =-4+w,\beta ^{10}=w,$ as we can
see from in below procedures:\medskip
\begin{verbatim}
A := -5*4^{-1}mod 61; for a to 61 do 
b := a^{10}mod 61; if b = 14 then print(a);fi;od;
   14 10 17  26 29 30 30 31 32 35 44 51                                 
 
 
for i from -31 to 30 do for j from -31 to 30 do
c :=(14*j+i) mod 61 d := (3/4)*j^2+(i+(1/2)*j)^2; 
if d < 61 and c = 10 then print(i, j)fi;od;od;
                                    -4, 1
                                    1, 5
                                    5, -4
 
\end{verbatim}

In this case, the rate code is $R_{1}=\frac{6k}{p-1}=\frac{k}{10},$ where $k$
is the dimension of the code, since we can't find $\beta $ such that $\beta
^{6}=w$ or $\beta ^{M_{j}}=w,$ for $M_{j}\mid p-1,j\in \{1,2,...,v\}.$

If \ \ we consider $p$ under the form $p=1+15\cdot 4,$ we use the real
Sedenion algebra, we get $n=4$ and for $w=\frac{1}{8}\left( 1+\overset{16}{%
\underset{i=2}{\sum }}e_{i}\right) ,$ we have $\pi =-1+16w,$ the label for $%
w $ is $s=42$ and $\beta =2+2w.$ In this case, the rate of the code is $%
R_{2}=\frac{15k}{p-1}=\frac{k}{4}$ and it is greater than $R_{1}.$ We remark
that we can use both algebras to define codes, but in the second case, we
have chance to obtain a better rate.(The dimension $k$ is considered the
same, in both situations).\medskip
\begin{verbatim}
 
A :=16^{-1}mod 61; for a to 61 do b :=a^{4}mod61; 
if b = 42 then print(a);fi;od;
42 25 30 31 36                                
 
 
for i from -31 to 30 do for j from -31 to 30 do c :=42*j+i mod 61;
 d := (15/64)*j^2+(i+(1/8)*j)^2; if d < 61 and c = 25 then print(i, j); 
fi;do;do;
                                   -6, 8
                                   -5, -8
                                    -2, 5
                                   -1, -11
                                    2, 2
                                   3, -14
                                    6, -1
 
\end{verbatim}

\textbf{Example 4.7. } Let \ $p=151=4+3\cdot 49=16+15\cdot 9=6\cdot 25+1.$

We have $t\in \{2,4\}$ and will use the real Quaternion algebra or real
Sedenion algebra. For $w=\frac{1}{2}\left( 1+\overset{4}{\underset{i=2}{\sum 
}}e_{i}\right) ,$ we have $\pi =-3+14w,n=25$ and $s=140,$ the label for $w.$
In this case, we \ can't find an element $\beta ,$ such that $\beta ^{25}=w,$
\ $\beta ^{6}=w,\beta ^{15}=w,$etc, as we can see in the procedure
below.\medskip
\begin{verbatim}
A:=-3*14^{-1} mod 151; for a to 151 do b:=a^25 mod 151; 
if b = 140 then print(a);fi;od:
                                     140
 
\end{verbatim}

But, as we remarked, the number $p$ can be written under the form $%
p=16+15\cdot 9=25\cdot 6+1,$ then if we take $t=4$, we can use the real
Sedenion algebra. We consider $w=\frac{1}{8}\left( 1+\overset{16}{\underset{%
i=2}{\sum }}e_{i}\right) .$ We obtain $\pi =1+24w,n=6$ and $s=44,$ the label
for $w.$ We can find $\beta ,$ such that $\beta ^{6}=w$ mod $\pi $ and $%
\beta =3-3w,$ with the label $s=22.\medskip $
\begin{verbatim}
A:=-24^{-1}mod 151; for a to 151 do 
b:=a^6 mod 151; if b = 44 then print(a);fi;do;
                  44 22 51 100 122 129
                                     
for i from -76 to 75 do for j from -76 to 75 do 
c := 44*j+i mod 151; d:= (15/64)*j^2+(i+(1/8)*j)^2;
 if d < 151 and c = 22 then print(i, j);fi;od;od;
                                   -9, 11
                                   -4, -20
                                    -3, 4
                                    3, -3
                                    4, 21
                                   9, -10
 
\end{verbatim}

\textbf{Example 4.8.} Let $p=149=25+\mathbf{31}\cdot 4=121+\mathbf{7}\cdot
4. $ In this situation, $t\in \{3,5\}$ \ and we can use the real Octonion
algebra or a real Cayley-Dickson algebra of dimension $32$.

We can't use the Algorithm 2 for octonions, since we can't obtain the
element $\beta $ and $p$ is not under the form $7k+1.$ For $w=\frac{1}{4}%
\left( 1+\overset{8}{\underset{i=2}{\sum }}e_{i}\right) ,$ we have $\pi
=9+8w.$ We consider $p=1+4\cdot 37$ and we can't find an element $\beta ,$
even if we take $p=2k+1$ or $4k+1$ or $37k+1.\medskip $
\begin{verbatim}
A := -9*8^{-1} mod 149; for a to 149 do 
b := a^2 mod 149; if b = 92 then print(a);fi;od;
                                     92
 
A := -9*8^{-1}mod 149; for a to 149 do
b := a^4 mod 149; if b = 92 then print(a);fi;od;
                                     92
 
A := -9*8^{-1}mod 149; for a to 149 do 
b := a^37 mod 149 if b = 92 then print(a);fi;od;
                                     92
                                     
\end{verbatim}

But we can choose another $\alpha .$ For example, for $w=\frac{1}{8}\left( 1+%
\overset{8}{\underset{i=2}{\sum }}e_{i}\right) ,$ we have $\pi =9+16w,$ and $%
s_{1}=46,~$the label for $w.$ If we consider $p=74n+1,~n=2,$ we get $\beta
=-2+4w,$ with label $s_{2}=33.$ In this case, the rate of the code is $R_{1}=%
\frac{74k}{p-1}=\frac{k}{2}.$ \ For $p=37n+1,$ $n=4,~$\ the label of $\beta
=4w$ is $s_{3}=35.$ In this case the rate of the code is $R_{2}=\frac{37k}{%
p-1}=\frac{k}{4}.$ We have $R_{2}<R_{1}.$ Therefore the code in the first
case is better, since it can have a greater rate as in the second case. For $%
p=2k+1$ or $4k+1,$ we can't find $\beta .\medskip \medskip $
\begin{verbatim}
A := -9*16^{-1} mod 149; for a to 149 do
 b := a^2 mod 149; if b = 46 then print(a);fi;od;
                              46  33 116
                                     
 
for i from -75 to 74 do for j from -75 to 74 do 
c := (46*j+i) mod 149); d := (31/64)*j^2+(i+(1/8)*j)^2; 
if d < 149 and c = 33 then print(i, j);fi;do;do;
                                   -4, 17
                                    -2, 4
                                    0, -9
                                    9, 7
                                   11, -6
 
A := -9*16^{-1} mod 149; for a to 149 do
b := a^4 mod 149; if b = 46 then print(a);fi;do;
               46 35 50 99 114
                                     
 
for i from -75 to 74 do for j from -75 to 74 do 
c :=(46*j+i) mod 149); d := (31/64)*j^2+(i+(1/8)*j)^2; 
if d < 149 and c = 35 then print(i, j);fi;do;do;
                                   -11, 1
                                   -2, 17
                                    0, 4
                                    2, -9
 
A := -9*16^{-1} mod 149; for a to 149 do 
b := a^37 mod 149; if b = 46 then print(a);fi;od;
                                     46
 
\end{verbatim}

If we work on a real algebra of dimension 32, let $w=\frac{1}{16}\left( 1+%
\overset{32}{\underset{i=2}{\sum }}e_{i}\right) .~$We have $\pi
=3+32w,s=107, $ the label for $w,\beta =4,$ with the label $%
s=4,n=4,p=37\cdot 4+1,$ as we can see in the procedures below.\medskip
\begin{verbatim}
 A := -3*32^{-1} mod 149; for a to 149 do
 b := a^4 mod 149; if b = 107 then print(a);fi;od;
                       107 4 27 122 145
                                      
 
for i from -75 to 74 do for j from -75 to 74 do
c := (107*j+i) mod 149; d := (31/256)*j^2+(i+(1/16)*j)^2; 
if d < 149 and c = 4 then print(i, j);fi;od;od;
                                   -8, 21
                                   -7, -18
                                   -4, 14
                                   -3, -25
                                    0, 7
                                   1, -32
                                    4, 0
                                    8, -7
                                   12, -14
 
 
\end{verbatim}

We \ can work on both algebras to obtain codes with good rate.\medskip

\textbf{Conclusions.} Regarding a finite field as a residue field modulo a
prime element from $\mathbb{V},$ where $\mathbb{V}$ is a subset of a real
algebra$\ $obtained by the Cayley-Dickson process with a commutative ring
structure, in this paper, we obtain an algorithm, called Main Algorithm,
which allows us to find codes with a good rate. This algorithm offers more
flexibility than other methods known until now, similar to Lenstra's
algorithm on elliptic curves compared with $p-1$\ Pollard's algorithm.

As a further research, we intend to improve this algorithm. This thing can
be done if \ first we can find answers at the following questions:

i) When the equation $\left( 4.1\right) $ has solutions?

ii) When we have more than one solution for the equation $\left( 4.1\right)
? $

iii) If we have solution(s) for equation $\left( 4.1\right) ,$ when we can
find the element $\beta ?$ 

\medskip

\textbf{References}%
\begin{equation*}
\end{equation*}%
[Co; 89] D. Cox, \textit{Primes of the Form} $x^{2}+ny^{2}$\textit{: Fermat,
Class Field Theory and Complex Multiplication}, A Wiley - Interscience
Publication, New York, 1989.\newline
[Da,Sa,Va; 03] G. Davidoff, P. Sarnak, A. Valette, \textit{Elementary Number
Theory, Group Theory, and Ramanujan Graphs}, Cambridge University Press,
2003.\newline
[Fl; 15] C. Flaut, \textit{Codes over a subset of Octonion Integers},
accepted in Results Math., DOI: 10.1007/s00025-015-0442-6.\newline
[Gh, Fr; 10] F. Ghaboussi, J. Freudenberger, \textit{Codes over Gaussian
integer rings}, 18th Telecommunications forum TELFOR 2010, 662-665.\newline
[Gu; 13] M. G\"{u}zeltepe, \textit{Codes over Hurwitz integers}, Discrete
Math., 313(5)(2013), 704-714.\newline
[Hu; 94] K. Huber, \textit{Codes over Gaussian integers}, IEEE Trans.
Inform. Theory, 40(1994), 207--216.\newline
[Ko, Mo, Ii, Ha, Ma; 10] H. Kostadinov, H. Morita, N. Iijima, A. J. Han
Vinck, N. Manev, \textit{Soft Decoding Of Integer Codes and Their
Application to Coded Modulation}, IEICE Trans. Fundamentals, E39A(7)(2010),
1363-1370.\newline
[Ma, Be, Ga; 09] C. Martinez, R. Beivide, E. Gabidulin, \textit{Perfect
codes from Cayley graphs over Lipschitz integers}, IEEE Trans. Inform.
Theory \textbf{55(8)(2009),} 3552--3562.\newline
[Ne, In, Fa, Pa; 01] T.P. da N. Neto, J.C. Interlando, M.O. Favareto, M.
Elia, R. Palazzo Jr., \textit{Lattice constellation and codes from quadratic
number fields}, IEEE Trans. Inform.Theory \textbf{47(4)(2001)} 1514--1527.%
\newline
[Ni, Hi; 08] S. Nishimura, T. Hiramatsu, \textit{A generalization of the Lee
distance and error correcting codes}, Discrete Appl Math., \textbf{156(2008)}%
, 588 -- 595.\newline
[Ri; 95] J. Rif\`{a}, \textit{Groups of Complex Integer Used as QAM Signals}%
, IEEE Transactions on Information Theory, \textbf{41(5)(1995)}, 1512-1517.%
\newline
[Sa; 14] D. Savin, \textit{Some central simple algebras which split}, An. 
\c{S}t. Univ. Ovidius Constan\c{t}a, \textbf{22(1)(2014)}, 263-272.\newline
[Si, Ta; 92] J. H. Silverman, J. T. Tate, \textit{Rational Points on
Elliptic Curves}, Springer-Verlag New York, 1992.

\medskip

Cristina FLAUT

{\small Faculty of Mathematics and Computer Science,}

{\small Ovidius University,}

{\small Bd. Mamaia 124, 900527, CONSTANTA,}

{\small ROMANIA}

{\small http://cristinaflaut.wikispaces.com/}

{\small http://www.univ-ovidius.ro/math/}

{\small e-mail:}

{\small cflaut@univ-ovidius.ro}

{\small cristina\_flaut@yahoo.com}

\end{document}